# Demonstration of new possibilities of multilayer technology on resistive microstrip/ microdot detectors

V. Cairo[a], R. De Oliveira[a], P. Fonte[b], S. Franchino[a], V. Peskov[a], P. Picchi[c], F. Pietropaolo[d]

[a] CERN, Geneva, Switzerland
[b] LIP, Coimbra, Portugal
[c] INFN Frascati, Italy
[d] INFN Padova, Italy

E-mail: vladimir.peskov@cern.ch

ABSTRACT: The first successful attempts to optimize the electric field in Resistive Microstrip Gas Chamber and resistive microdot detectors using additional "field shaping" strips located inside the detector substrate are described

*Corresponding author

# Contents

## 1. Introduction

Gaseous micropattern detectors open a new page in gaseous detector developments, allowing photons and charged particles to be recorded with unprecedented high position resolution: up to 10-40 μm. An important modification has been the implementation of resistive electrodes in some of the designs [ 1-3 ] which has made the latter quite robust and also spark-protected.
Several years ago a multilayer printed circuit technology, widely used nowadays in microelectronics, was applied to micropattern detector design and manufacturing (see [3-5]). First it was implemented in micropattern detectors with resistive electrodes. These modified detectors contain additional strips or other shape electrodes located either under the resistive electrodes [4] or inside the detector substrate [5] and were used to detect either the charge or the induced signals produced by avalanches in the amplification gap. Such a double layer structure allows the avalanche position to be determined with a high accuracy and, of course, ensures full protection of the front end electronics, connected to the inner strips, preventing it from being damaged in the case of occasional sparks.

The aim of the work was to investigate another possible functionality of inner strips: electric field tuning in a double layered resistive micropattern detectors by applying voltages to these strips. This can be useful in some cases, for example, when the surface streamers limit the maximum achievable gas gains, as usually happens in microstrip gas chambers (MSGC) [6, 7]. In the latter case an effective measure is to minimize the field parallel to the surface. For this reason resistive microstrip gaseous chambers (RMSGCs) and resistive microdot detectors were chosen for our studies. These detectors are promising in RICH and dual-phase noble liquid applications [8]. Our team was especially interested in further improvements of the resistive microdot detector in order to incorporate them into the new design on a dual- phase noble liquid detector containing a CsI photocathode immersed inside the liquid [3]. The aim is to reach 100% efficiency in detecting single photoelectrons. The microdot detector should operate at gas gains more than $3\times10^{4}$.

## 2. First tests with RMSGC having double layered electrodes

The most straightforward way for us to test the possible influence of voltage- biased inner strips on the electric field in the avalanche region was to use RMSGC-based design. RMSGC is a relatively simple device which was systematically studied by us earlier [8]. It could be easily modified by adding inner strips of any geometry.
The first prototype used in our preliminary studies (see [9]) is shown in Figure 1.

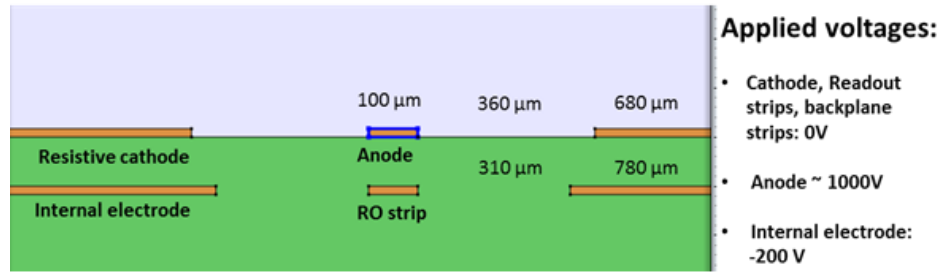

**Figure 1**. MSGC geometry and applied voltages used for the tests and electrostatic simulations.

It is a double layer RMSCG having anode and cathode strip widths of 100 μm and 680 μm respectively and a pitch of 1.5 mm. At 100 μm below these, inside the substrate, an additional array of strips is located oriented parallel to the top electrodes. The strips under the anodes were used to detect the avalanche-induced signals whereas the strips under the cathode can be biased by the voltage and therefore were used for the electric field optimization in the avalanche gap.
Before making measurements simulations (COMSOL Multiphysics) were performed to clarify the expected field modifications for various voltages applied to the inner electrodes located under the cathode strips. As examples, results of simulations for two voltage settings are shown in Fig. 2:
1) 1100 V applied to the anode strip with all other electrodes grounded (Fig. 2a)
2) in addition, -200 V is applied to the electrodes located under the cathode strips (Fig. 2b).
Note that in both cases the drift field was 1.5 kV/cm.

In Figs. 3a,b results of calculations of the electric field strength along the surface in the direction perpendicular to the strips are shown. As can be seen from Fig. 3a, when the inner electrodes were grounded, the electric field on the MSGC surface had a high value not only in the anode region, but also near the cathode edges. This effect is more pronounced in the case of the standard MSGC, without embedded electrodes (see [9]).

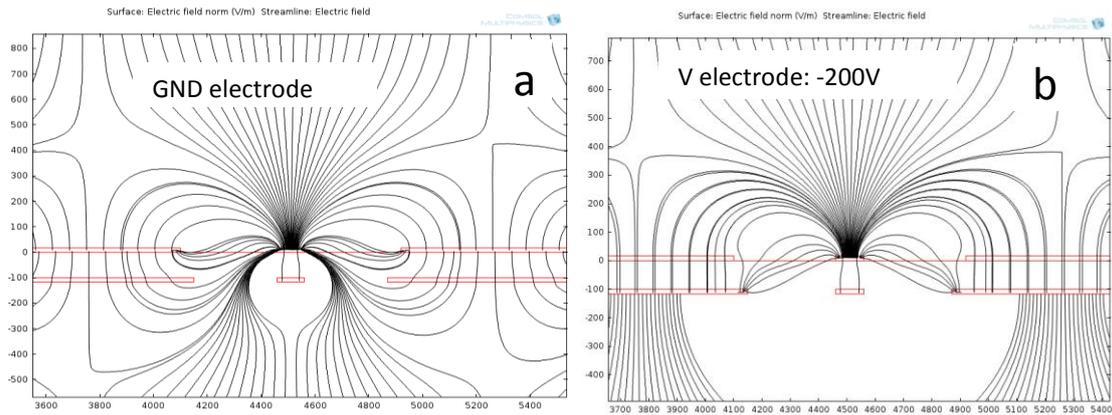

**Figure 2**. Simulated field lines in the device: a) the case when 1100 V is applied to the anode strip, whereas all other electrodes are grounded. b) the case when additionally -200V is applied to the inner electrodes located under the cathode strips.

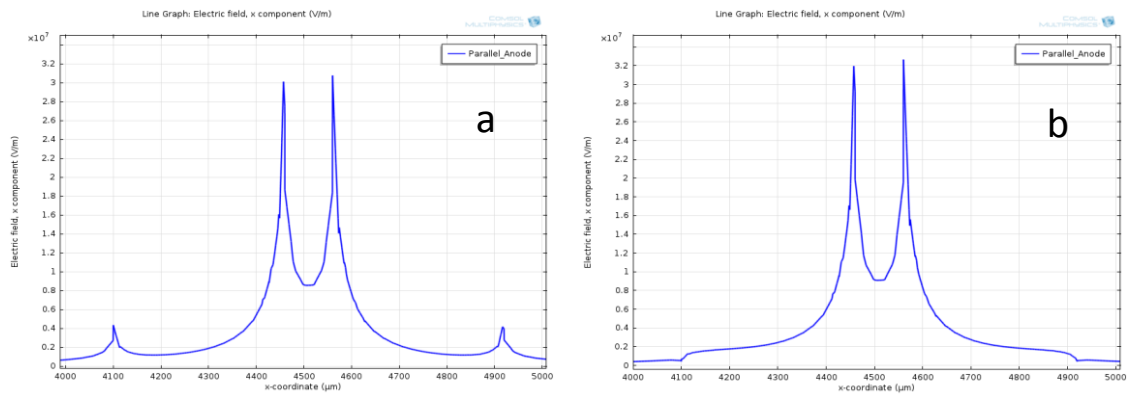

**Figure 3**. Electric field scan in a direction perpendicular to the strips and parallel to the surface: a) grounded internal electrodes; b) -200V applied to the inner cathode strips. Two large peaks in the center of the figure correspond to the electric field in the anode region. Two other peaks are visible at the cathode edges in the case in when internal strips are grounded.

Dramatic changes happened when -200V were applied to the inner electrodes (Fig 3b): in this case the electric field near the anode strips increased, whereas the peaks near the cathode strips almost disappeared. Such field modifications may lead to an increase of the maximum achievable gain of the MSGC and a more stable detector operation. Indeed, as was shown in [6,7], usually surface streamers limit the maximum achievable gain in MSGCs. The favourable factors for their formation are:

1) High value of electric field near the edges, where some streamers can originate

2) High enough electric field component along the substrate surface necessary to support streamer propagation.

Of course, as was mentioned earlier, these calculations are purely electrostatic and in a real situation there could be a significant contribution from avalanche ions attached on the substrate surface and modifying the electric field.

In any case, in practice, the calculated field line maps cannot easily be verified. However, one can try to find some indirect indications. For example, the maximum achievable gain as a function of the voltages applied to the inner strips can be measured and checked whether it really increases.

This was the aim of the experimental part of this work. For the detector studies the same experimental setup was used as in earlier works (see [8, 9]). The tests were performed in Ne, Ar and their mixtures with 20-30% of $CO_2$. As radioactive sources $^{241}$Am or $^{55}$Fe were used. The signals can be detected from anode strips or from the readout strips located under the anode. Various measurements were performed; as an example, in Fig. 4 results of RMSGC gain measurements performed at various applied voltages to the inner cathode strips $V_{el}$ with a detector with 30 μm anode strips are shown. It is clearly seen that the gas gain increases with $V_{el}$ reaching values close to $10^5$ which is almost ten times higher than those achieved earlier with an ordinary RMSGC [8] having a similar geometry. This indicates that inner electrodes can indeed be used for the improvement of the detector performance. More results obtained with this detector one can find in [9].

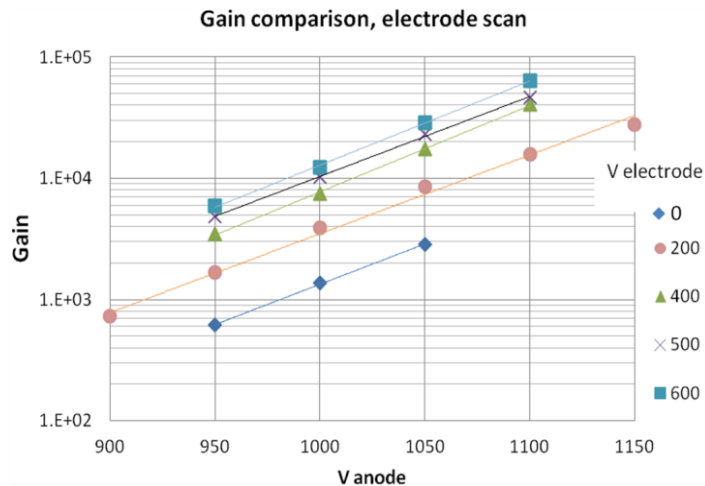

**Figure 4**. Gas gains of RMSGC vs. the anode voltage measured at different negative $V_{el.}$ Detector with 30 μm strips, used gas: Ar+30% $CO_2$.

## 3. Resistive microdot detectors

After obtaining validation of the principle with an RMSGC-based design, we tried to implement the same concept to much more sophisticated device-resistive microdot detectors. As was already mentioned in the introduction, microdot detectors are the subject of our main interest in connection with their possible application in a new dual-phase noble liquid detector described in [3].

Note that earlier prototypes of resistive microdot detectors (let's call them "version 0") were manufactured from MSGCs by coating anode strips with a resistive layer having small circular opening [3]. In geometry field lines do not have azimuthal symmetry around the opening since most of the field lines are oriented perpendicular to the anode and the cathode strips. However, preliminary results obtained with resistive microdot detectors were quite encouraging [3].

In this work we made two significant modifications to the resistive microdot detectors:

1) Spiral-shape anodes were introduced instead of strip used in version 0. This made the field around the dots more uniform in the azimuthal direction which, as will be shown later, allowed the maximum achievable gas gain to be increased 2-3 times;

2) Inner cathode electrodes were added allowing the maximum achievable gain to be boosted by another factor of 10, so the total improvement in the gain, compared to version 0 was a factor of 20-30.

Consequently we will first describe the microdot detector with resistive spiral anodes (let's call them "version 1") and then we will compare this design and results obtained with the more advanced version having both spiral anodes and inner cathode strips ("version 2"). We kept the same anode-cathode gaps and the same applied voltages in order to better perform the comparison.

**3.1 Detectors design and manufacturing steps**

Both versions of those microdot detectors were manufactured at the printed circuits workshop at CERN as a standard multilayer Printed Circuit Board (PCB) using the photolithographic technique. The layout of both prototypes is shown in Fig 5.

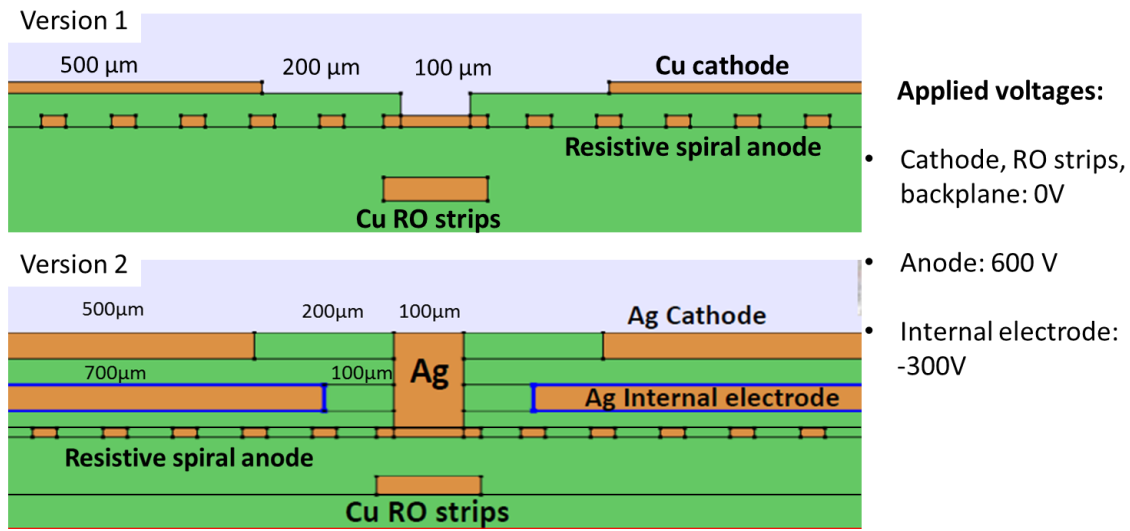

**Figure 5**. Layout of the cross section of both versions of resistive spiral detectors. Top: First version, bottom: second one. Voltages used for the simulation and tests are also indicated.

**3.1.1 Manufacturing of version 1**

The manufacturing process can be divided into two phases: the anode board construction and the cathode plane. All the production steps are summarized in Fig 6.

The anode board manufacturing started with a 2.4 mm thick fiber glass plate (FR4, EMC 370) with 35 µm of copper on both sides. On the bottom of it, the entire layer of Cu was preserved, in order to act as a grounded backplane of the final detector. On the top of it, readout strips (150 µm wide) were chemically etched (Fig 6 a).

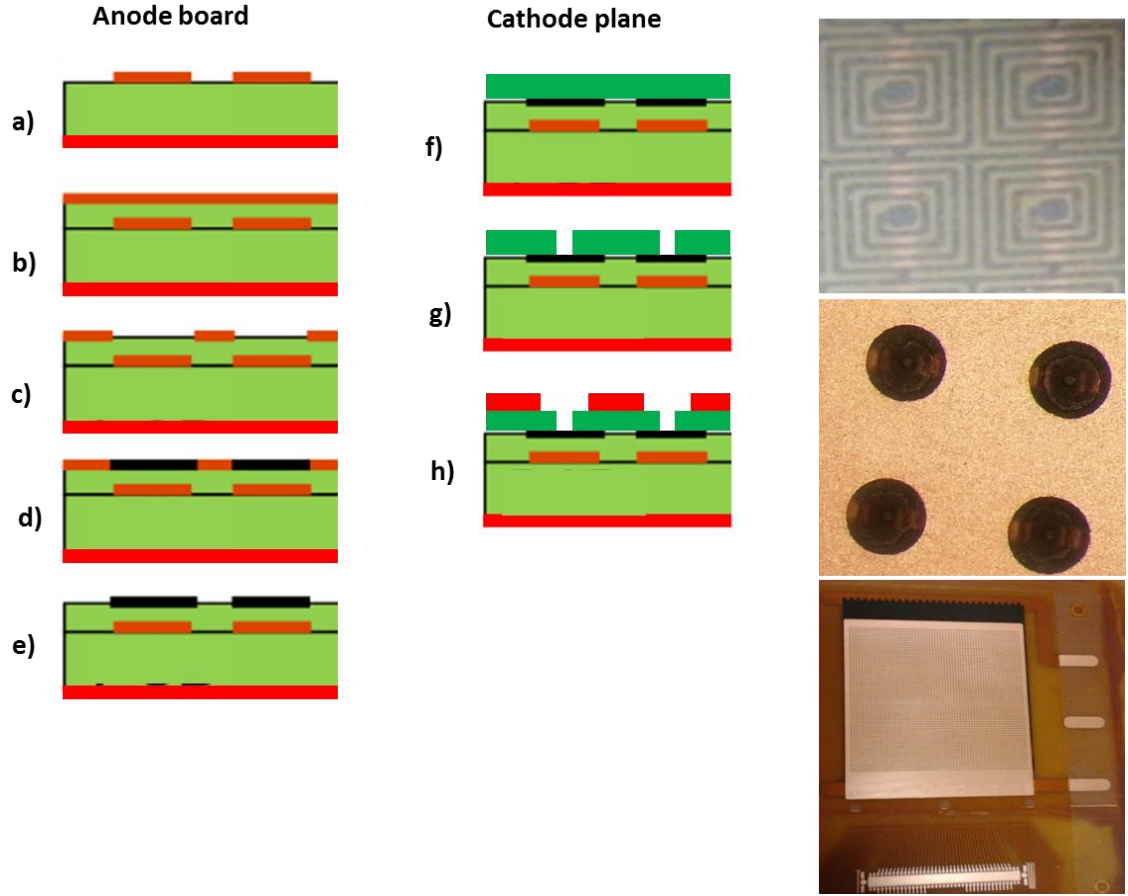

**Figure 6**. Manufacturing steps schematics (left part for the anode board and centre one for cathode plane and anode openings). On the right are some pictures taken during the construction. Top: resistive anode spirals (in black), aligned with readout strips (in yellow). Centre: the cathode copper plane and the anode micro-holes. Bottom: the final detector.

A 75 µm thick dielectric (glass fiber and epoxy glue) and 17 µm of copper were then pressed on top of the first PCB layer (Fig. 6b). The role of the dielectric is to decouple the induced signal readout strips from the anode area of the detector, where the avalanche takes place.

Then resistive anodes were created on the copper layer. First, spiral-shape grooves, 35 µm wide, were etched onto this copper layer (Fig. 6c), and then filled with resistive paste with surface resistivity of 1 MΩ/□ (ESL RS 12116), as shown in Fig. 6d. Once the spiral image was created, all the remaining copper was etched away (Fig. 6e).

In order to produce the cathode plane, a dielectric material (Pyralux Photoimageable coverlay by DuPont, 50 µm thick) was laminated on top of the PCB anode board (Fig. 6f) and cylindrical dots, of diameter 100 µm and aligned with the center of the spirals, were created on it using a photolithographic technique (Fig. 6g).

A layer of previously drilled (holes of diameter 500 μm) copper was then pressed and glued on top of the anode board, paying attention to the alignment of the apertures (Fig. 6h). This layer acted as a cathode plane.

On the right of Fig. 6, some pictures taken during the manufacturing are shown: on top are the resistive spirals, in the centre the cathode plane and anode micro-holes and at the bottom the final detector.

The detector pitch was 1 mm and the total active area was 6*6 cm$^2$. The resistive value was measured to be around 4 GΩ between the centre of the spiral anodes and the connector that provides voltage to all the resistive layer.

### 3.1.2 Manufacturing of version 2

The second version of the spiral detector was recently produced with the implementation of some improvements in production and with the introduction of a corrective electrode layer, as already suggested earlier (see the layout at the bottom of Fig.5). The pitch (1 mm) and microhole diameter (100 μm) were kept the same as in the first version.

The manufacturing improvements with respect to version 1 are:

1) The use of photoimageable coverlay (Pyralux Photoimageable coverlay by DuPont) as dielectric and as a mask for all electrode planes (anode, corrective electrodes and cathode).

2) The use of silver glue (Polytec EC 101) as conductive material for the internal electrode and cathode plane. In this way the pressing process of the cathode plane is avoided, improving the relative alignment between anode dots and cathode rings.

The other advantage of this geometry is the flat surface, avoiding problems of dust during detector operation.

The first step was to produce a readout plane with parallel readout strips, as in the first version (Fig. 6a).

In addition, as can be seen from Fig 5 bottom, the detector is made with the other six layers created with Coverlay as dielectric material and silver paste as conductor (apart from the spiral anode plane that is created with the resistive paste). These pastes fill the complementary image created with photolithography in the coverlay. Fig 7 shows the layout of each layer and the picture of the result. In the layout schematics, black indicates where the coverlay remains after having created the image. The grooves created in the coverlay that will be filled by the paste that creates each electrode are in white.

The first coverlay layer (64 μm thick) has been laminated on top of the readout strips. In this case no grooves were created in it in order to ensure insulation between capacity coupling strips and active area on top of them.

The next layers were all 50 μm thick coverlay. In the second layer grooves with the complementary image of spirals were created (Fig 7 left). Then this image was filled with resistive paste (ESL RS 12116) in order to create anode spirals (in black in Fig 7 centre).

On top of those spirals, the third layer was deposited and dots of 100 μm, corresponding to the centre of the spirals, were made. These dots were filled with conductive silver glue in order to bring the electric contact between the centre of spiral anodes to the surface of the detector.

On top of that, the fourth coverlay layer was deposited in which the complementary image of the internal electrodes was grooved (Fig 7 centre) and the corresponding image was filled again with conductive silver paste in order to create the conductive electrode plane.

The last two layers again consisted of the layout of 100μm dots, (the electric contact of the anode) and the last one with the cathode plane plus the micro-holes connected to the center of the anode spirals (Fig 7 right).

Every layer, after deposit of the coverlay and the creation of the image with photolithographic techniques, was cured in an oven at 180°C in order to strengthen the coverlay or the deposited paste. In the case of the paste, after baking, it was polished with sandpaper to produce a flat surface, suitable for the creation of the next layer.

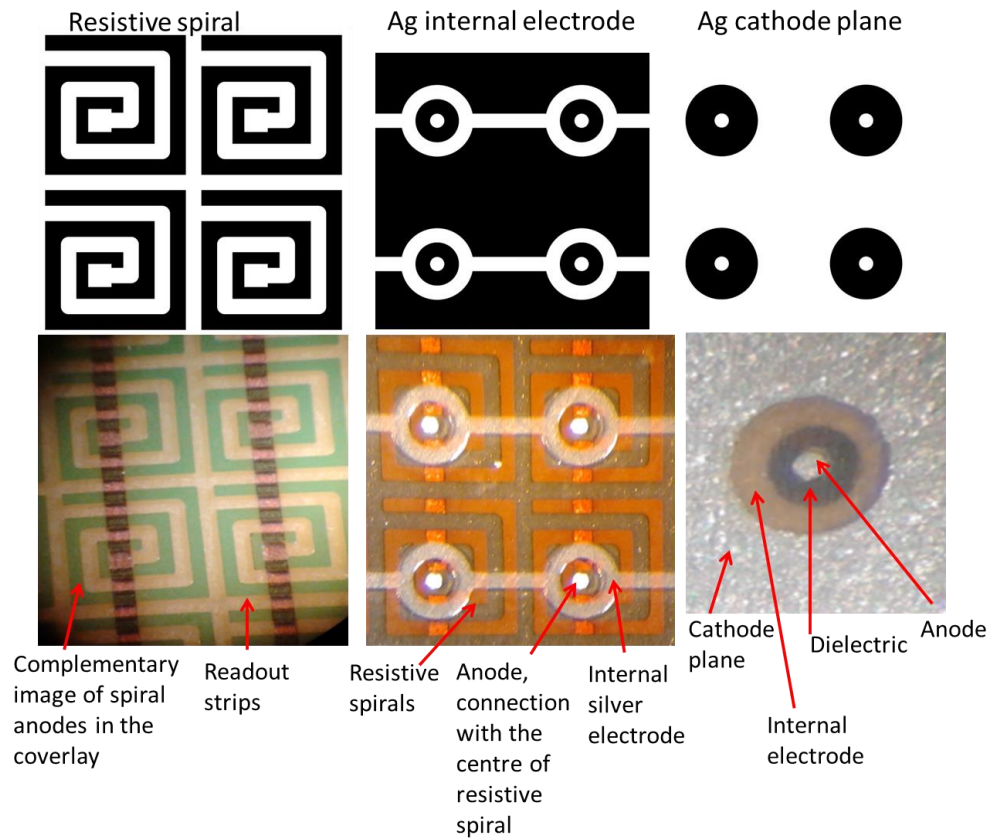

**Figure 7.** Top row: layout of the negative image, created in the coverlay, of the electrodes on top of readout strips. The resistive spiral image is filled with resistive paste, while the internal electrode and the cathode are created with the conductive silver glue. Bottom row: magnified pictures of the result of each electrode layer.

The chosen technique has the advantage of being able to monitor the alignment of each layer during the manufacturing and before curing the materials. In this way a particular layer can be recreated if any serious defects appear on it. Every layer was also tested with high voltage before the depositing of the next one.

Fig. 7 bottom left shows the grooves with the spiral image created in the coverlay before curing. The alignment between the spiral centre and the readout strips can be seen.

In the bottom centre picture the readout strips, resistive spirals, internal silver electrodes and microdots, all centred on the resistive spiral centre are clearly visible.

**3.2 Simulation**

We have simulated electric fields in bothe devices (version 1 and 2) using COMSOL Multiphysics with the electrostatic package.

The main goal was to optimize the electric field strength distribution, trying to avoid peaks outside the active area which could create instabilities and sparks. We also tried to improve the collection of field lines at the center of the anode, avoiding field-lines entering into the material, that could create an undesirable charging up effect from avalanche ions accumulating in this region. As in areal detector we have introduced the internal electrode into the geometry with rings concentric to the anode and embedded in the material, below the cathode. The voltage on the internal ring was set around -300V.

The result of the simulation of both versions are shown in Fig 8, in terms of electric field-lines (left) and the intensity of electric field on a line parallel to the surface (right).

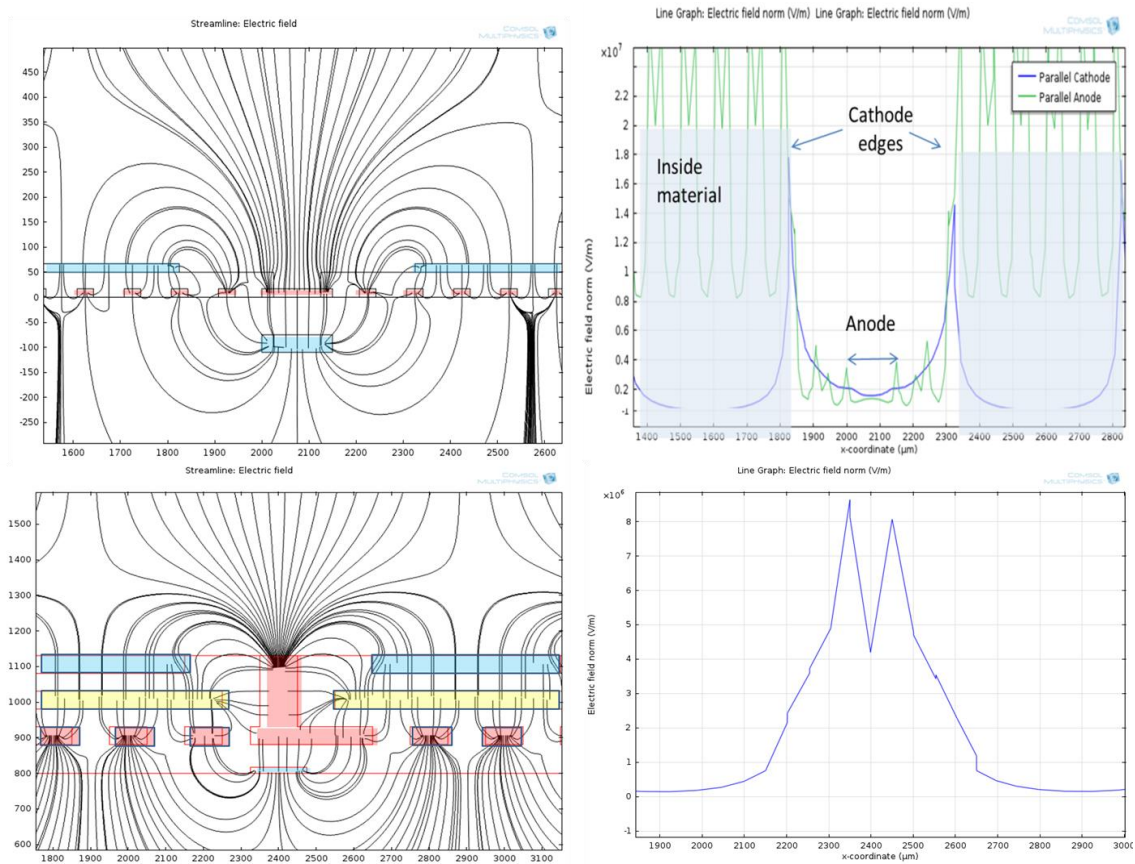

**Figure 8**. Results of simulation. Top row: first version, bottom row- second one. Left: electric field lines, right electric field strength versus the coordinate along the surface in the direction perpendicular to the anode dots row. In the figures on the left side, the electrodes are highlighted in different colours depending on the applied voltages: cathode and readout strips, at ground are in blue, the resistive strips at 600V are in red and the corrective electrode at -300V is in yellow.

In the case of the first version, the anode and cathode do not lay on the same plane, so the electric field is plotted both for the direction parallel to the cathode (blue) and to the anode (green). In this case many peaks appear but most of them correspond to the edges of the spiral and they are hidden inside the material, while the electric field at the edges of the cathode is higher than the one at the edges of the anode, where the avalanche takes place. As already pointed out earlier, this could cause the formation of streamers.

Note that in the second version this is no longer the case, thanks to the internal corrective electrodes, as was already shown in the MSGC case (Fig 3b, [9]). The other advantage of having the electrodes is that in the second version we reduced the number of electric field-lines entering in the material, hopefully reducing charging up effects.

## 4. Results on resistive microdot detectors

### 4.1 Version 1

Gain curves measured with a microdot detector having spiral anodes (version 1) and operating in Ne or Ar are shown in Fig. 9. As can be seen gas gains $(4-5) \times 10^4$ were achieved which are 2-3 times higher than were obtained with the resistive microdot detector version 0 [3]. The same maximum achievable gains, but at higher operating voltages was obtained in a mixture with $CO_2$, indicating that the quencher concentration does not play any important role in this particular detector geometry.

The energy resolution as a function of gas gain is shown in Fig. 9 right. This dependence has a minimum (~40%) in a gas gain interval of 2000-3000. At lower gains the energy resolution was spoiled due to the contribution from the amplifier noise. At higher gains the energy resolution degraded due to the avalanche statistics.

As in the case of any detector with resistive electrodes, the gas gain of microdot detector version 1 falls with the counting rate as shown in Fig. 10. The chosen resistive value for the electrodes has not yet been optimized in order to cope with higher rates.

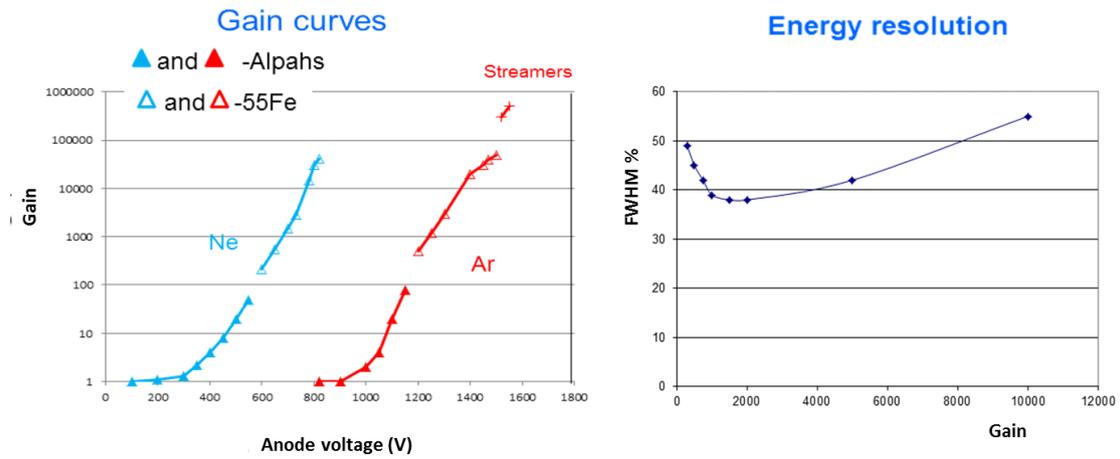

**Figure 9.** Results of the resistive microdot detector, version 1 in terms of gain versus anode voltage for Ne and Ar gases (left) and $^{55}$Fe energy resolution as a function of a gas gain measured in Ar.

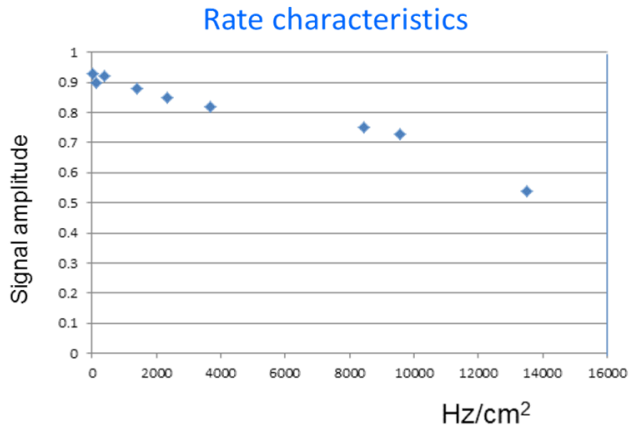

**Figure 10.** Rate characteristics of a microdot detector version 1.

**4.2 Version 2**

The main results obtained with microdot detector version 2 operating in Ar+30%$CO_2$ gas mixture are presented in Fig. 11-12. In Fig. 11 the dependence of the gas gain on the anode voltage is depicted, when -325 V were applied to its inner electrodes. In this case the maximum achievable gas gain is approaching $(3-4) \times 10^5$ which is ten time higher than that achieved with microdot detector version 1.

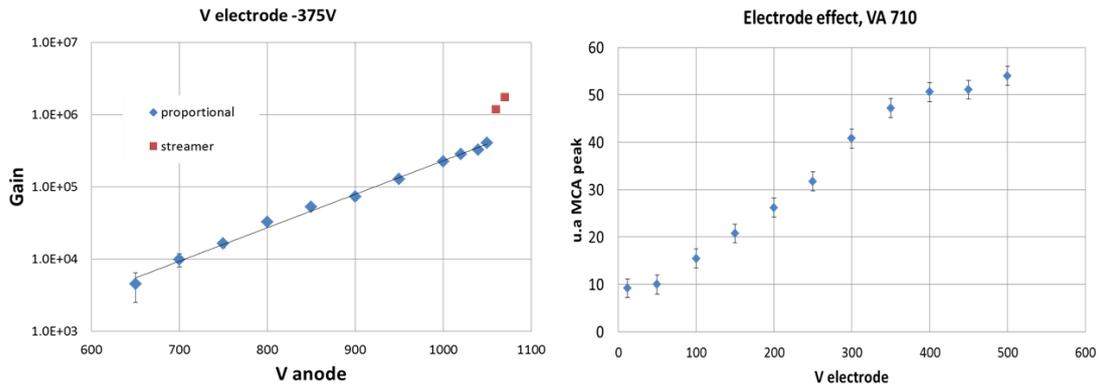

**Figure 11**. Left: gain curve as a function of the anode voltage for an electrode voltage of -375V. Right: electrode effect on the gain at anode voltage of 710 V.

The typical pulse-height spectrum of the $^{55}$Fe measured at two anode voltages (720, 740 V) is shown in Fig. 12 (left). The dependence of the energy resolution on the anode voltage is shown in Fig. 12 (right). As in the case of the microdot detector version 1, at low anode voltages (low gas gains) the energy resolution degraded due to the amplifier noise.

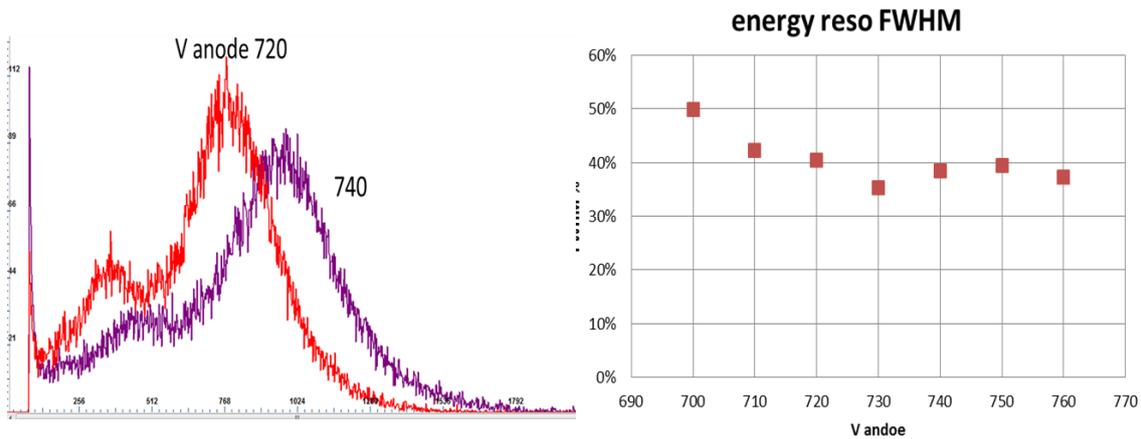

**Figure 12**: Left: MCA spectrum of $^{55}$Fe source with two anode voltages (720, 740 V), right: energy resolution (FWHM) measured as a function of the anode voltages.

### 4.3. The time stability studies of the spiral detector

We have also investigated the time stability properties of the spiral detector, making a distinction between effects due to the application of high voltage (dielectric polarization effect) and the ones that are source dependent (radiation effect).
The voltage effect is presumably due to the slow movement of ions inside the dielectric material to the new equilibrium state, once the electric field is applied across it. This usually gives a detector gain increase as a function of time with a plateau reached after a certain time depending on the dielectric material .
The radiation effect (a charging up effect caused by the $^{55}$Fe source ) is due to the accumulation of charges on the dielectric surfaces which affect the electric field in the active area of the detector, causing a change of the gas gain. This effect depends on the gain and on the rate of the source.

### 4.3.1. Voltage effects

For these measurements we used a $^{55}$Fe source fixed at a given position, in order to avoid space-dependent effects. A shutter was used to open or close the source. To minimize the radiation effect each energy spectrum was taken with the source on for 10s, and then the shutter was closed untill the start of the next measurement.
The voltage effects were studied with the floating inner electrode as well as when -375V was applied to it. In both cases the anode voltage was kept at the same value, 710V. In the case of the floating electrode the initial gain was about 5000, in the case of the powered electrode it was about 20000. In Fig. 13 (left) a typical resultis shown : the peak position of the MCA spectrum as a function of the time at which the voltage was applied to the detector. In both cases the gain is increasing, however, in the case of the floating electrode the total gain reached a plateau after ~30 min with a gain increase on a factor of ~ 2 whereas in the case of the powered electrode, the plateau was reached only after a few minutes and the total gain increase effect was only 30% .

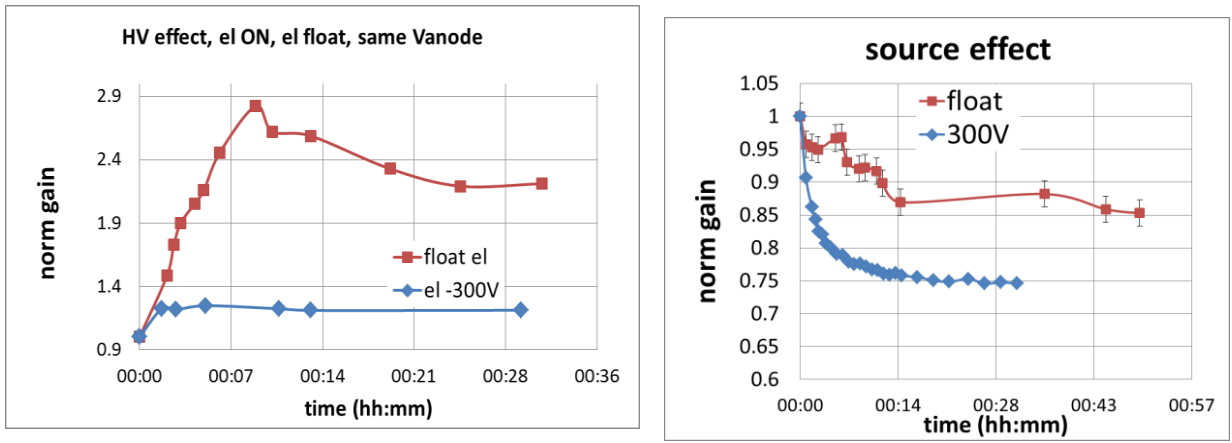

**Figure 13**. Gain stability versus time (hours: min) in the case of floating electrode (red curve) and powered electrode (blue curve). Left: effect of voltage; zero at x-axis corresponds to the time when the voltage was applied to the detector. Right: effect of the source. In this case the $^{55}$Fe source was activated (time zero) after the detector was stabilized due to the voltage effect.

### 4.3.2. Radiation effects

In order to investigate the effects of the radiation on the gas gain (charging up effects) we proceeded in the following way. We left the $^{55}$Fe source (a counting rate was~ 2kHz/cm$^2$) open in the same place of the detector, but only after a plateau was reached corresponding to the voltage effect. The obtained result can be seen in Fig. 13 right, for both floating (red) and powered electrode (blue). The time zero in this plot corresponds to the time when the source shutter was open. The detector stabilizes after ~ 15 minutes, in both cases at -15% gain loss in the case of the floating electrode and -25% gain loss in the case of the powered electrode. In the case of the powered electrode the effect is a little bigger, probably due to the higher gain.
We have also investigated the repeatability of the previous measurements. The results can be seen in Fig. 14 in the case of the powered electrode. We found out that both effects are reproducible. In the right-hand part of the figure one can see the total effect of voltage and source. The result is a rapid increase of gain, due to the voltage and a slow decrease due to the charging up. The total effect of the gain reduction is only about 5%.

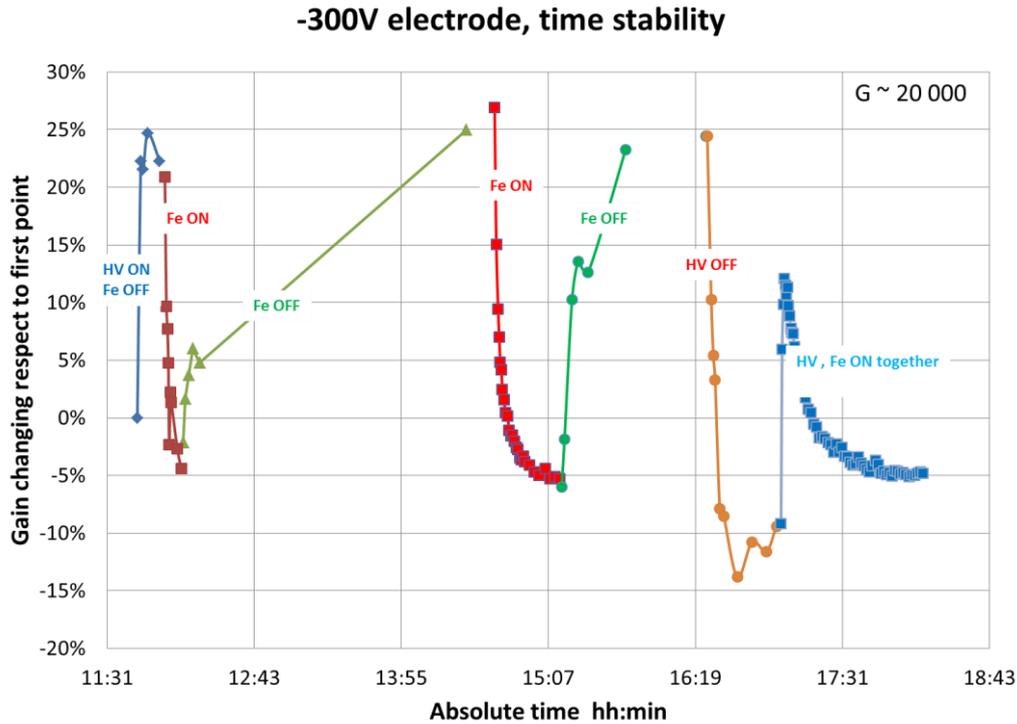

**Figure 14**: Studies of the combined effect of the applied voltage and radiation on the detector gain electrode. variation. In these measurements 710V was applied to the anode and -300V to the internal

Note that this value obtained of 5% is quite good if compared to the stability vs. the time observed at various conditions with GEM or TGEM [10-12].

## 5. Discussion and future plans

Preliminary results obtained with double-layered RMSGC indicate that the effect of inner voltage-biased strips on the maximum achievable gain was clearly seen, at least at the low counting rates used in our work.
Additionally the filed line map was significantly optimized in microdot detectors with spiral anodes allowing higher gains to be achieved and allowing measurements to be carried out in a region far away from where instabilities start to be seen.

It must be noted that in these preliminary tests the intermediate layer between the detector active electrodes and the inner strips was not optimized: at present it is made of fiber glass FR4 for the first version and coverlay for the second one, which have too high a resistivity to prevent the detector from operating consistently at high counting rates. Our previous experience with thick GEMs (TGEMs) made of fiber glass show that two effects may contribute towards the detector instability:

    1) a long-term polarization effect of the dielectric layers,
    2) charging up the surface by positive ions

The first effect is well known and was observed both with GEMs and TGEMs [10-12]. The simplest solution is to keep a constant high voltage applied to the detector [13]. Then after some time (1-3 hours) the detector gain is stabilized.

The second effect leads to a dynamic situation: ions deposited on the surface can temporally change the local electric field as it happens, for example, in resistive plate chambers (RPC). However after some time:

$$\tau \sim \varepsilon_0 \varepsilon \rho \quad (1),$$

(where $\varepsilon_0$ and $\varepsilon$ are the permittivity of the vacuum and the intermediate layer between the active electrode and the inner strips respectively and $\rho$ is the latter resistivity). When the charge dissipates, the field line map will return to the initial one. Hence one can expect that the inner electrodes can efficiently influence the field only at a counting rate below some critical value $1/\tau$.

To minimize both effects, in the next prototype the intermediate layer between upper and lower strips will be made from another material having a much lower (~$10^{10}\Omega$cm) resistivity. We are also designing prototypes in which the top dielectric layer is made of a special resistive material [14], the resistivity of which can be regulated by the manufacturing procedure.

Note that attempts to influence the field near the anode of the MSGC were made quite a long time ago. For example some designs of the MSGC had a so-called back plane to which the voltage was applied [15]. An MSGC having anode and cathode strips on the opposite sides of the detector substrate was also tested [16]. There were some indications that the inner strips in resistive MICROMEGAS under certain condition also affect the detector gain [17].

All these early studies and our recent observations show that extra electrodes located under or close to the active electrodes can indeed be used for the tuning of electric fields in micro pattern detectors. This may offer new possibilities in their design and optimization.

# 6. Conclusions and future plans

In this work it was demonstrated for the first time that by biasing the inner electrode strips in a multilayer micro pattern detector (RMSGC and resistive microdot) one can essentially modify the electric field and obtain much higher gas gains than in usual micropattern detectors. For example, gas gains of $(3-4) \times 10^5$ were achieved with a microd detector having resistive spiral anode and field correction strips. This makes the latter attractive for many applications especially for those which deal with single electron detection.

Our plans now are to optimize the material of the electrodes and their resistivity in order to make microdot detectors suitable for cryogenic applications. We also consider developing a spiral detector having hexagonal shape of electrodes, as it was suggested by S. Biagi [18]. Of course, the final tests will be performed at cryogenic temperatures

# Acknowledgments

We would like to warmly thank E. Oliveri, L. Ropelewski for frequent discussions and help throughout this work.